\documentclass[onecolumn,showpacs,showkeys,preprintnumbers]{revtex4}
\usepackage{amsthm, amscd, amsfonts, amssymb, graphicx}
\usepackage{dcolumn}
\usepackage{bm}
\usepackage[english]{babel}

\begin{document}

\title{The model of a hypothetical room-temperature superconductor.}
\author{Konstantin V. Grigorishin}
\email{gkonst@ukr.net}
\author{Bohdan I. Lev}
\email{bohdan.lev@gmail.com}
\affiliation{Boholyubov Institute for
Theoretical Physics of the National Academy of Sciences of
Ukraine, 14-b Metrolohichna str. Kiev-03680, Ukraine.}
\date{\today}

\begin{abstract}
The model of hypothetical superconductivity, where the energy gap
asymptotically approaches zero as temperature increases, has been
proposed. Formally the critical temperature of such a
superconductor is equal to infinity. For practical realization of
the hypothesis a superconducting material with such properties is
predicted.
\end{abstract}


 \maketitle

\section{Introduction}\label{intr}

A main problem of technical application of superconductors
consists in that their critical temperatures $T_{\texttt{c}}$ are
considerably lower than room temperature. The critical temperature
depends on an effective coupling with some collective excitations
$g=\nu_{F}\lambda$ (here $\nu_{F}$ is a density of states at Fermi
level, $\lambda$ is an interaction constant) and on energy of the
collective excitations $\omega$. Most often a phonon mechanism
results in superconductivity, then $g$ is an electron-phonon
coupling constant (in most cases $g\lesssim 1$), and $\omega$ is a
characteristic phonon frequency $\omega\sim 100\div 400\texttt{K}$
(here $\hbar=k_{B}=1$). The larger coupling constant, the larger
critical temperature. At large $g$ (as a rule for $g>10$) we have
following expressions for the critical temperature
\cite{mahan,ginz}:
\begin{eqnarray}
  T_{\texttt{C}} &\propto& \omega g\quad-\quad \texttt{BCS} \quad \texttt{theory}  \label{1.1}\\
  T_{\texttt{C}} &\propto& \omega\sqrt{g} \quad-\quad \texttt{Eliashberg} \quad \texttt{theory} \label{1.2}
\end{eqnarray}
Formally the critical temperature can be made arbitrarily large by
increasing the electron-phonon coupling constant
$T_{\texttt{c}}(g\rightarrow\infty)\rightarrow\infty$. However, in
order to reach room temperature such values of the coupling
constant are necessary, which are not possible in real materials.
Moreover we can increase the frequency $\omega$ due nonphonon
pairing mechanisms, as proposed in \cite{ginz}. However with
increasing of the frequency the coupling constant decreases as
$g\propto 1/\omega$, therefore
\begin{equation}\label{1.3}
T_{\texttt{c}}(\omega\rightarrow\infty)=1.14\omega\exp\left(-\frac{1}{g}\right)\rightarrow
0.
\end{equation}
Many different types of superconducting materials with a wide
variety of electron pairing mechanisms exist, however all they
have the critical temperature limited by values $\lesssim
100\texttt{K}$, despite the fact that the highly exotic mechanisms
have been proposed. In a present work we propose a fundamentally
different approach to the problem of room-temperature
superconductivity. This approach is not associated with increasing
of the coupling constant or with change of the frequency, but it
allows to circumvent the problem in the sense that, having the
interaction of conventional intensity (which generates an energy
gap $\Delta\sim 10\div 100\texttt{K}$), we change ratio between
the gap and the critical temperature as
$2\Delta/T_{\texttt{c}}\rightarrow 0$ instead of a finite value
$3\div 7$ for presently known materials.

\section{General idea.}\label{general}
First of all we propose a principal possibility to increase the
critical temperature due to generalization of BCS model in the
following sense: let us consider a system of fermions with
Hamiltonian:
\begin{eqnarray}\label{2.1}
    \widehat{H}=\sum_{\textbf{k},\sigma}\xi(k)a_{\textbf{k},\sigma}^{+}a_{\textbf{k},\sigma}
    -\frac{\lambda}{V}\sum_{\textbf{k},\textbf{p}}a_{\textbf{p}\uparrow}^{+}a_{-\textbf{p}\downarrow}^{+}a_{-\textbf{k}\downarrow}a_{\textbf{k}\uparrow}
    +\upsilon\sum_{\textbf{k}}\left[\frac{\Delta}{|\Delta|}a_{\textbf{k}\uparrow}^{+}a_{-\textbf{k}\downarrow}^{+}
    +\frac{\Delta^{+}}{|\Delta|}a_{-\textbf{k}\downarrow}a_{\textbf{k}\uparrow}\right]
    \equiv \widehat{H}_{\texttt{BCS}}+\widehat{H}_{\texttt{ext}},
\end{eqnarray}
where $\widehat{H}_{\texttt{BCS}}$ is BCS Hamiltonian - kinetic
energy + pairing interaction ($\lambda>0$), energy $\xi(k)\approx
v_{F}(|\textbf{k}|-k_{F})$ is counted from Fermy surface. The term
$\widehat{H}_{\texttt{ext}}$ is the external pair potential or
"source term" \cite{matt1}. For example, in ferromagnetism a term
$\textbf{SH}$ - the energy of a spin in an external magnetic field
plays a role the source term $\widehat{H}_{\texttt{ext}}$.
Operators
$a_{\textbf{k}\uparrow}^{+}a_{-\textbf{k}\downarrow}^{+}$ and
$a_{-\textbf{k}\downarrow}a_{\textbf{k}\uparrow}$ are creation and
annihilation of Cooper pair operators \cite{schr}, $\Delta$ and
$\Delta^{+}$ are anomalous averages:
\begin{eqnarray}\label{2.2}
    \Delta^{+}=\frac{\lambda}{V}\sum_{\textbf{p}}\left\langle
    a_{\textbf{p}\uparrow}^{+}a_{-\textbf{p}\downarrow}^{+}\right\rangle,
    \quad
    \Delta=\frac{\lambda}{V}\sum_{\textbf{p}}\left\langle
    a_{-\textbf{p}\downarrow}a_{\textbf{p}\uparrow}\right\rangle,
\end{eqnarray}
which are the complex order parameter $\Delta=|\Delta|e^{i\theta}$
(the energy gap $\Delta$ is analogous to magnetization
$\textbf{M}=\langle \textbf{S} \rangle$ in ferromagnetism). The
multipliers $\frac{\Delta}{|\Delta|}$ and
$\frac{\Delta^{+}}{|\Delta|}$ are introduced into
$\widehat{H}_{\texttt{ext}}$ in order that the energy does not
depend on the phase $\theta$ ($a\rightarrow
ae^{i\theta/2},a^{+}\rightarrow a^{+}e^{-i\theta/2}\Longrightarrow
\Delta\rightarrow\Delta e^{i\theta},\Delta^{+}\rightarrow
\Delta^{+}e^{-i\theta}$). Thus both $\widehat{H}_{\texttt{BCS}}$
and $\widehat{H}_{\texttt{ext}}$ is invariant under the $U(1)$
transformation unlike the source term in \cite{matt1} where it has
a noninvariant form
$\upsilon\sum\left[a_{\textbf{k}\uparrow}^{+}a_{-\textbf{k}\downarrow}^{+}
+a_{-\textbf{k}\downarrow}a_{\textbf{k}\uparrow}\right]$. Hence
$\upsilon$ is energy of a Cooper pair relative to uncoupled state
of the electrons in the external pair potential
$H_{\texttt{ext}}$. It should be noted that the energy gap
$|\Delta|$ is energy of a Cooper pair relative to uncoupled state
of the electrons too. However the field $\Delta$ is a
self-consistent field as a consequence of attraction between
electrons. The field $\upsilon$ is the applied field to the system
from the outside.

\begin{figure}[ht]
\includegraphics[width=8.0cm]{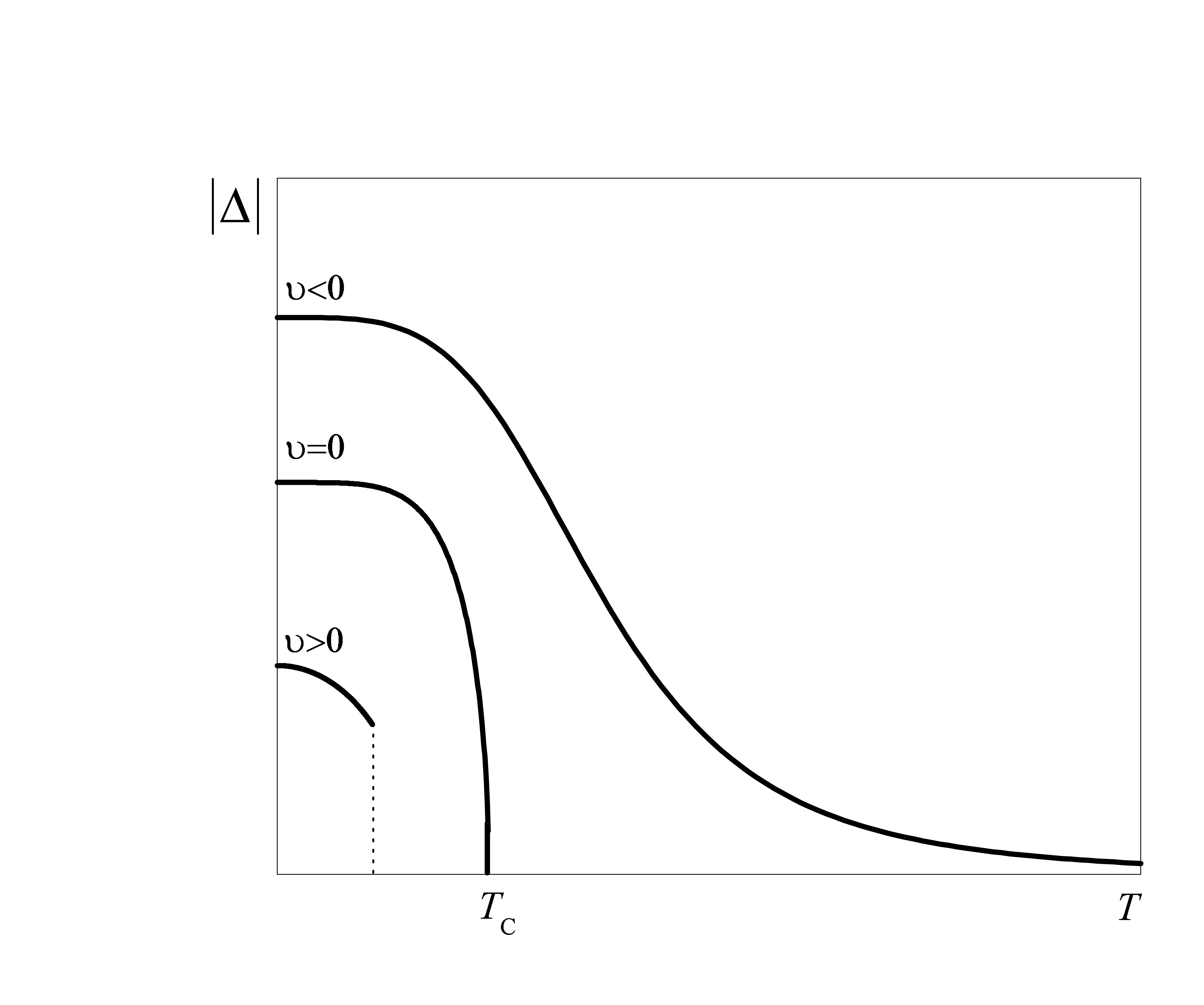}
\caption{Energy gaps $\Delta(T)$ as solution of Eq.(\ref{2.6}) for
three values of the external pair potential $\upsilon$.}
\label{Fig1}
\end{figure}

Using the Fermi commutation relations and the anomalous averages
(\ref{2.2}), the Hamiltonian (\ref{2.1}) can be rewritten in a
form \cite{tinh}:
\begin{eqnarray}\label{2.3}
    \widehat{H}=\sum_{\textbf{k},\sigma}\xi(k)a_{\textbf{k},\sigma}^{+}a_{\textbf{k},\sigma}
    +\left(1-\frac{\upsilon}{|\Delta|}\right)\sum_{\textbf{k}}\left[\Delta^{+}a_{\textbf{k}\uparrow}a_{-\textbf{k}\downarrow}
    +\Delta a_{-\textbf{k}\downarrow}^{+}a_{\textbf{k}\uparrow}^{+}\right]+\frac{1}{\lambda}V|\Delta|^{2}.
\end{eqnarray}
Then normal $G$ and anomalous $F$ propagators have forms:
\begin{eqnarray}
    G=i\frac{i\varepsilon_{n}+\xi}
    {(i\varepsilon_{n})^{2}-\xi^{2}-|\Delta|^{2}(1-\upsilon/|\Delta|)^{2}}\label{2.5a}\\
    F=i\frac{\Delta(1-\upsilon/|\Delta|)}
    {(i\varepsilon_{n})^{2}-\xi^{2}-|\Delta|^{2}(1-\upsilon/|\Delta|)^{2}},\label{2.5b}
\end{eqnarray}
where $\varepsilon_{n}=\pi T(2n+1)$ \cite{matt2}. Then from
Eq.(\ref{2.2}) we have self-consistency condition for the order
parameter
\begin{equation}\label{2.6}
    \Delta=\lambda \nu_{F}
    T\sum_{n=-\infty}^{\infty}\int_{-\omega}^{\omega}d\xi iF(\varepsilon_{n},\xi)
    \Longrightarrow 1=g\int_{-\omega}^{\omega}d\xi
    \frac{1-\upsilon/|\Delta|}{2\sqrt{\xi^{2}+|\Delta|^{2}(1-\upsilon/|\Delta|)^{2}}}
    \tanh\frac{\sqrt{\xi^{2}+|\Delta|^{2}(1-\upsilon/|\Delta|)^{2}}}{2T}.
\end{equation}
Solutions of Eq.(\ref{2.6}) are shown in Fig.{\ref{Fig1}}. If the
external pair potential is absent $\upsilon=0$ we have usual
self-consistency equation for the gap $\Delta$: the gap is a
function of temperature such that $\Delta(T\geq
T_{\texttt{c}})=0$. The larger coupling constant
$g=\lambda\nu_{F}$, the larger $T_{\texttt{c}}$. If $\upsilon>0$
then the pairing of quasiparticles results in increase of the
system's energy that suppresses superconductivity and first order
phase transition takes place. If $\upsilon<0$ then the pairing
results in decrease of the system's energy. In this case a
solution of Eq.(\ref{2.6}) is such that the gap $\Delta$ does not
vanish at any temperature. At large temperature
$T\gg\omega,|\upsilon|$ the gap is
\begin{equation}\label{2.7}
    |\Delta(T\rightarrow\infty)|=\frac{g\omega|\upsilon|}{2T}.
\end{equation}
Hence the critical temperature is $T_{\texttt{c}}=\infty$ (in
reality it limited by the melting of the substance). In
ferromagnetism if the external magnetic field $\textbf{H}$
presents then the magnetization exists at any temperature and
$\textbf{M}(T\rightarrow\infty)\rightarrow 0$. It should be noted
that if $\lambda=0$ then for any $\upsilon$ a superconducting
state does not exist ($\Delta=0$ always). This means
electron-electron coupling is the cause of the transition to
superconducting state only but not the external pair potential
$\upsilon$. This fact is a peculiarity of superconductivity and it
does not have analogous in ferromagnetism. The theory of
superconductivity with Hamiltonian (\ref{2.1}) has been developed
in a work \cite{grig3} similarly to Ginzburg-Landau theory, the
effect of a Coulomb pseudopotential has been investigated in a
work \cite{grig4}.

Let us take into account the fact that electron-phonon interaction
$gD(\omega(\textbf{q}))$ leads to superconductivity, where $D$ is
a phonon propagator. Eliashberg equations \cite{mahan,ginz},
unlike BCS equations, can describe renormalization of
quasiparticles' mass and decrease of effectiveness of the
interaction at phonon energies $\omega(\textbf{q})\lesssim T$ -
difference of the asymptotics (\ref{1.1}) and (\ref{1.2}). However
these facts does not change the previous conclusions. In
\cite{levit,grig1,grig2} a method to consider electron-phonon
interaction has been proposed, however the resulting equation is
much simpler than Eliashberg equations. Phonons are dispersionless
$\omega(q)=\omega$ and the electron-phonon coupling constant does
not depend on a wave vector $\lambda(q)=\lambda$ are suggested in
this method. The electron-phonon interaction generates the gap via
the anomalous propagator $F$ and renormalizes an energetic
parameter via the normal propagator $G$:
\begin{eqnarray}
\Delta_{n}&=&\int\frac{d^{3}p}{(2\pi)^{3}}\lambda^{2}
T\sum_{m=-\infty}^{+\infty}iF(\textbf{p},\varepsilon_{m})iD(\varepsilon_{n}-\varepsilon_{m},\textbf{k})\nonumber\\
&=&g \sum_{m=-\infty}^{+\infty} \frac{\pi
T\Delta_{m}}{\sqrt{\widetilde{\varepsilon}_{m}^{2}+|\Delta_{m}|^{2}}}
\frac{\omega^{2}}{(\varepsilon_{n}-\varepsilon_{m})^{2}+\omega^{2}}
\label{2.9}\\
\widetilde{\varepsilon}_{n}&=&i\varepsilon_{n}+\int\frac{d^{3}p}{(2\pi)^{3}}\lambda^{2}
T\sum_{m=-\infty}^{+\infty}iG(\textbf{p},\widetilde{\varepsilon}_{m})iD(\varepsilon_{n}-\varepsilon_{m},\textbf{k})\nonumber\\
&=&i\varepsilon_{n}+g\sum_{m=-\infty}^{+\infty} \frac{\pi
Ti\widetilde{\varepsilon}_{m}}{\sqrt{\widetilde{\varepsilon}_{m}^{2}+|\Delta_{m}|^{2}}}
\frac{\omega^{2}}{(\varepsilon_{n}-\varepsilon_{m})^{2}+\omega^{2}},\label{2.10}
\end{eqnarray}
where $g\equiv\lambda^{2}\nu_{F}\frac{2}{\omega_{0}}$ is the
coupling constant. The phonon propagator can be represented in a
form:
\begin{equation}\label{2.11}
    \frac{\omega^{2}}{(\varepsilon_{n}-\varepsilon_{m})^{2}+\omega^{2}}\rightarrow
    \frac{\omega}{\sqrt{\varepsilon_{n}^{2}+\omega^{2}}}\frac{\omega}{\sqrt{\varepsilon_{m}^{2}+\omega^{2}}}
    \equiv w_{n}w_{m},
\end{equation}
and the gap depends on the energetic parameter as
$\Delta_{n}=\Delta w_{n}$. Then the energetic parameter is not
renormalized: $\widetilde{\varepsilon}_{n}=i\varepsilon_{n}$,
because there is an odd function under the sign of sum in
(\ref{2.10}) in the approximation (\ref{2.11}). Equation for the
gap has a form:
\begin{eqnarray}\label{2.12}
1=g \sum_{n=-\infty}^{+\infty} \frac{\pi
Tw^{2}_{n}}{\sqrt{\varepsilon_{n}^{2}+|\Delta|^{2}w^{2}_{n}}}
\end{eqnarray}
When $T_{\texttt{c}}\ll\omega$ (small $g$) Eq.(\ref{2.12}) has the
asymptotic (\ref{1.3}), when $T_{\texttt{c}}\gg\omega$ (large $g$)
one has the asymptotic (\ref{1.2}). Eq.(\ref{2.12}) is easier than
Eliashberg equations, however one is more correct than BCS
equation.

Substituting the propagators (\ref{2.5a},\ref{2.5b}) in
Eqs.(\ref{2.9},\ref{2.10}) we have an analog of Eq. (\ref{2.6}):
\begin{eqnarray}\label{2.13}
1=g \sum_{n=-\infty}^{+\infty} \frac{\pi
T\left(w_{n}-\upsilon/|\Delta|\right)w_{n}}{\sqrt{\varepsilon_{n}^{2}+|\Delta|^{2}(w_{n}-\upsilon/|\Delta|)^{2}}}
\end{eqnarray}
If to suppose $\upsilon=0$ then Eq.(\ref{2.13}) is transformed
into Eq.(\ref{2.12}). The function $\Delta(T)$ for the cases
$\upsilon>0,\upsilon=0,\upsilon<0$ is the same as in
Fig.(\ref{Fig1}). If $\upsilon<0$ then in a limit
$T\gg\omega,|\upsilon|$ we have
\begin{equation}\label{2.14}
    |\Delta(T\rightarrow\infty)|=\frac{2g\omega|\upsilon|}{\pi T}.
\end{equation}
that is analogous to Eq.(\ref{2.7}).

\section{Model of a superconductor.}\label{model}

In the previous section we demonstrated a principal possibility to
increase the critical temperature due to the external pairing
potential $\upsilon$. In this section we propose a model of the
system where such a situation can be realized. Let us consider
superconductors alkali-doped fullerenes
$\texttt{A}_{3}\texttt{C}_{60}$
($\texttt{A}=\texttt{K},\texttt{Rb},\texttt{Cs}$). The threefold
degenerate $t_{1u}$ level is partly occupied and electrons couple
strongly to eight $H_{g}$ intramolecular Jahn-Teller phonons
(electron-vibron interaction). Hamiltonian of the system has a
form \cite{han1}:
\begin{eqnarray}\label{3.1b}
\widehat{H}&=&-\sum_{ijm\sigma}t_{ij}a^{+}_{im\sigma}a_{jm\sigma}+U\sum_{imm'}n_{im\uparrow}n_{im'\downarrow}+\omega\sum_{i\nu}b^{+}_{i\nu}b_{i\nu}\nonumber\\
&+&\lambda\sum_{imm'\sigma\nu}V^{(\nu)}_{mm'}a^{+}_{im\sigma}a_{im'\sigma}\left(b^{+}_{i\nu}+b_{i\nu}\right).
\end{eqnarray}
$a^{+}_{im\sigma}(a_{jm\sigma})$ is the electron creation
(annihilation) operator acting on site $i$, orbital $m=1,2,3$ and
spin $\sigma$. $b^{+}_{i\nu}(b_{i\nu})$ is the phonon creation
(annihilation) operator with the vibration mode $\nu=1,\ldots,5$.
$t_{ij}$ is the hopping integral, $U$ is the on-site Coulomb
interaction, $\omega$ is the phonon frequency, and $\lambda$ is
the electron-phonon coupling constant. The coupling matrices
$V^{(\nu)}$ are determined by icosahedral symmetry. The
dimensionless electron-phonon coupling constant is
$g=\frac{5}{3}\lambda^{2}\nu_{\texttt{F}}/\omega$. Typical
parameters are $g\sim 0.5\div 1$, $\omega/W\sim 0.1\div 0.25$ and
$U/W\sim 1.5\div 2.5$, where $W\sim 0.5\texttt{eV}$ is a electron
bandwidth. Basic mechanisms resulting in superconductivity are:
\begin{enumerate}
    \item The dynamical Jahn-Teller effect (interaction with $H_{g}$
intramolecular oscillations) favors the formation of a local
singlet \cite{lam,lam2,han2}:
\begin{equation}\label{3.1}
    |\Phi_{0\uparrow\downarrow}\rangle=\frac{1}{\sqrt{3}}\sum_{m}C_{m\uparrow}^{+}C_{m\downarrow}^{+}|0\rangle,
\end{equation}
where the spin-up and spin-down electrons have the same $m$
quantum number, i.e., a local pairing takes place. Here
$|0\rangle$ is the neutral $\texttt{C}_{60}$ molecule for the
alkali-metal-doped materials, the quantum number $m$ labels the
three orthogonal states of $t_{1u}$ symmetry (LUMO state). The
normal state (high spin state) of two electrons is
\begin{equation}\label{3.1a}
    |\Phi_{0\uparrow\uparrow}\rangle=C_{m_{1}\uparrow}^{+}C_{m_{2}\uparrow}^{+}|0\rangle.
\end{equation}
The low-spin state is lower in energy than the high-spin state if
$E_{\texttt{JT}}>\frac{2}{3}K$, where
$E_{\texttt{JT}}=\frac{\lambda^{2}}{\omega}$ is a Jahn-Teller
energy, $K$ is an exchange integral. For
$\texttt{A}_{x}\texttt{C}_{60}$ ($x=2,3,4$) the coupling with
$H_{g}$ phonons overpowers Hund's rule coupling. Experimental
confirmation of this fact is that $\texttt{A}_{4}\texttt{C}_{60}$
must be anti-ferromagnetic insulator (Hubbard-like model
predicts), while it is known experimentally there are no moments
in $\texttt{A}_{4}\texttt{C}_{60}$.
    \item In a work \cite{han1} the following result has been obtained.
For noninteracting electrons the hopping tends to distribute the
electrons randomly over the molecular levels. This makes more
difficult to add or remove an electron pair with the same $m$
quantum numbers. However as $U$ is large $U>W$ the electron
hopping is suppressed and the local pair formation becomes more
important. The Coulomb interaction actually helps local pairing.
For $A_{g}$ phonons, the phonon-induced attractive interaction
$U_{\texttt{ph}}$ is of the order of $U_{\texttt{ph}}/W\sim
-0.47$. $T_{\texttt{C}}$ is vanished when $U+U_{\texttt{ph}}\geq
0$. For the Jahn-Teller $H_{g}$ phonons the attractive interaction
is smaller $U_{\texttt{ph}}/W\sim -0.2$. This attractive
interaction is, therefore, quickly overwhelmed by the Coulomb
repulsion. Superconductivity remains, however, even for
$U+U_{\texttt{ph}}\gg 0$, and $T_{\texttt{c}}$ drops surprisingly
slowly as $U$ is increased. The reason is that local pairing
arises from correlation of spin and orbital structures within each
site, and therefore it is not suppressed by the charge
interaction. Superconductivity is expected to exist in this
material right up to the Mott transition.

This situation is very different from Eliashberg theory. We can
see, because of the local pairing, the Coulomb interaction enters
very differently for Jahn-Teller and non-Jahn-Teller models, and
it cannot be easily described by a Coulomb pseudopotential:
$g-\mu^{\ast}$.
\end{enumerate}

\begin{figure}[h]
\includegraphics[width=7.0cm]{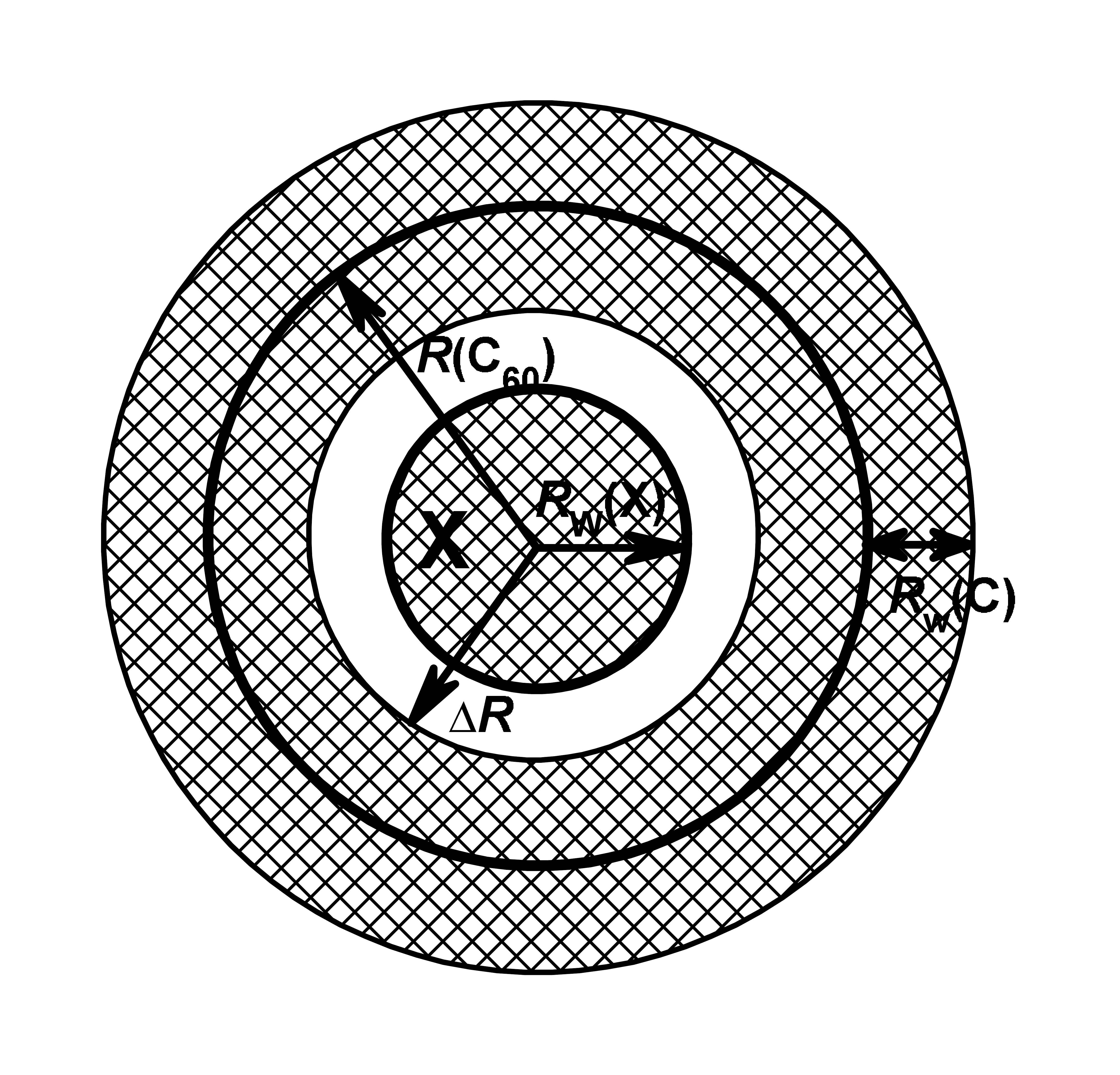}
\caption{Cross-section of an endohedral fullerene
$\texttt{X@C}_{60}$. The carbon cage can be considered as a
spherical layer of thickness $2R_{\texttt{W}}(\texttt{C})$ and
central radius $R(\texttt{C}_{60})$. The central atom $\texttt{X}$
placed into the inner cavity radius of $\Delta R$ is a noble gas
atom van der Waals radius of $R_{\texttt{W}}(\texttt{X})$.}
\label{Fig2}
\end{figure}

Thus in an alkali-doped fullerene the Cooper pairs are formed on
one molecule size of $R=3.55\texttt{A}$ as result a of
electron-vibron interaction and suppression of hopping between
molecules with one-cite Coulomb interaction $U$. This situation is
fundamentally different from superconductivity in metals, where
the size of a Cooper pair is macroscopic quantity $\sim 10^{3}\div
10^{4}\texttt{A}$. Let us consider some features of the molecular
structure $\texttt{C}_{60}$. The van der Waals radius of a carbon
atom is $R_{\texttt{W}}(\texttt{C})=1.70\texttt{A}$. Thus a
fullerene has an inner cavity in its center the size of $\Delta
R=R(\texttt{C}_{60})-R_{\texttt{W}}(\texttt{C})=1.85\texttt{A}$. A
noble gas atom $\texttt{X}$ can be trapped in a carbon cage in the
inner cavity (Fig.\ref{Fig2}) - we have endohedral complexes
$\texttt{X@C}_{60}$
\cite{breton,cios,saund,rubin,dennis,jime,patchk}. Since for a
helium atom $R_{\texttt{W}}(\texttt{He})=1.40\texttt{A}<\Delta R$
the atom interacts with the carbon cage by van der Waals
interaction only and it's electronic shell does not make
hybridized orbitals with electronic shells of the carbon cage. If
a helium atom is placed into each fullerene molecule in
alkali-doped fullerenes then we have hypothetical material
$\texttt{A}_{3}\texttt{He@C}_{60}$. Electronic properties of
$\texttt{A}_{3}\texttt{He@C}_{60}$ must be exactly the same as
electronic properties of $\texttt{A}_{3}\texttt{C}_{60}$. Changes
in oscillation spectrum of a fullerene can be neglected.

As noted above, in an endohedral fullerene the noble gas atom
interacts with a carbon cage by van der Waals force. As is well
known van der Waals interaction depends on electronic
configuration of interacting subsystems. In alkali-doped
fullerenes alkali metal atoms give valent electrons to fullerene
molecules. Then energy of the van der Waals interaction has to
depend on a state of the excess electrons on the surface of a
molecule $\texttt{C}_{60}$. Any two electrons can be in the paired
state (\ref{3.1}) or in the normal state (\ref{3.1a}). Let this
energy for the paired state is $\upsilon_{\uparrow\downarrow}$ and
the energy for the normal state is $\upsilon_{\uparrow\uparrow}$.
If $\upsilon_{\uparrow\downarrow}<\upsilon_{\uparrow\uparrow}$
then the paired state is more energetically favorable than the
normal state: a molecule $\texttt{X}@\texttt{C}_{60}^{n-}$ has
lower energy if the excess electrons are in the paired state than
the energy if the electrons are in the normal state (if we turn
off the electron-electron interaction). On the other hand as noted
above for noninteracting electrons the hopping $t_{ij}$ tends to
distribute the electrons randomly over the molecular levels thus
destroying the local pairs. However the relationship
$\upsilon_{\uparrow\downarrow}<\upsilon_{\uparrow\uparrow}$ makes
more energetically favorable to place electrons in a state
(\ref{3.1}) with the same quantum numbers $m$ and thus it
confronts the destruction of local pairs by the hopping. Hence a
function
\begin{equation}\label{3.2b}
    \upsilon=\upsilon_{\uparrow\downarrow}-\upsilon_{\uparrow\uparrow}
\end{equation}
plays the role of the external pair potential. The van der Waals
interaction is interaction due to virtual transitions of the
Cooper pair from a triply degenerated level $t_{1u}$ ($l=5$) to
levels $t_{1g}$ ($l=5$), $h_{g},t_{2u},h_{u}$ ($l=6$),
$g_{g},g_{u},t_{g}$ ($l=7$) \cite{allon,hadd}, where $l$ is an
orbital index for $\pi$-electrons (we can use a state
$C_{m\uparrow}^{+}C_{m\downarrow}^{+}|0\rangle$ instead of the
state (\ref{3.1}) for simplicity, because the result will not
depend on the quantum number $m$):
\begin{equation}\label{3.2}
    \Phi_{0\uparrow\downarrow}\equiv
    \Omega_{l=5,\gamma}(\textbf{R}_{1})\Omega_{l=5,\gamma}(\textbf{R}_{2})
    \longleftrightarrow
    \frac{1}{\sqrt{2}}\left[\Omega_{l,\gamma'}(\textbf{R}_{1})\Omega_{l=5,\gamma}(\textbf{R}_{2})
    +\Omega_{l=5,\gamma}(\textbf{R}_{1})\Omega_{l,\gamma'}(\textbf{R}_{2})\right]\equiv\Phi_{k\uparrow\downarrow},
\end{equation}
and of the helium atom from a level $1s$ to levels
$2s,2p,3s,3p,3d\ldots$:
\begin{eqnarray}\label{3.3}
    &&\Psi_{0}\equiv
    f_{0,0}(r_{1})Y_{0,0}(\textbf{r}_{1})f_{0,0}(r_{2})Y_{0,0}(\textbf{r}_{2})\nonumber\\
    &&\longleftrightarrow
    \frac{1}{\sqrt{2}}\left[f_{n,\widetilde{l}}(r_{1})Y_{\widetilde{l},\widetilde{m}}(\textbf{r}_{1})f_{0,0}(r_{2})Y_{0,0}(\textbf{r}_{2})
+f_{0,0}(r_{1})Y_{0,0}(\textbf{r}_{1})f_{n,\widetilde{l}}(r_{2})Y_{\widetilde{l},\widetilde{m}}(\textbf{r}_{2})\right]\equiv\Psi_{p},
\end{eqnarray}
where the index $\gamma$ labels irreducible representation of
icosahedral symmetry group; $n,l,m$ are principal quantum number,
orbital quantum number and magnetic quantum number accordingly;
$f_{n,l}(r)$ is a radial wave function, $Y_{l,m}$ is a spherical
wave functions. $\Phi_{0}$ and $\Psi_{0}$ are ground-states of a
Cooper pair and a helium atom accordingly. $\Phi_{k}$ and
$\Psi_{p}$ are the excited states of the Cooper pair and the
helium atom accordingly, $k$ and $p$ are sets of quantum indices
of the corresponding exited states. $\textbf{R}_{1}$ and
$\textbf{R}_{2}$ are radius-vectors of electrons of the Cooper
pair, and $|\textbf{R}_{1}|\approx|\textbf{R}_{2}|\approx R$ since
the Cooper pair is on surface of the molecule. $\textbf{r}_{1}$
and $\textbf{r}_{2}$ are radius-vectors of electrons of the helium
atom, and $\langle r\rangle=0.31\texttt{A}\ll R=3.55\texttt{A}$ -
the atom is much less than the fullerene molecule. Signs "+" in
the sums are caused by the fact that the ground states of both the
Cooper pair and the helium atom are singlet, and transitions
between singlet and triplet states are allowed only when there is
the spin-orbit interaction, but this interaction can be neglected.
Energy of the van der Waals interaction is defined with the second
order correction:
\begin{equation}\label{3.4}
    \upsilon_{\uparrow\downarrow}=\sum_{k,p}\frac{|\langle \Phi_{k\uparrow\downarrow},\Psi_{p}|\widehat{V}|\Psi_{0},\Phi_{0\uparrow\downarrow}\rangle|^{2}}
    {E_{0}+\widetilde{E}_{0}-E_{k}-\widetilde{E}_{p}},
\end{equation}
where the summation is made over the indexes of all possible
excited states of the Cooper pair $k$ and of the helium atom $p$;
$E_{0}$ and $\widetilde{E}_{0}$ are ground state energies of the
Cooper pair and the helium atom accordingly, $E_{k}$ and
$\widetilde{E}_{p}$ are energies of the corresponding excited
states. Since $E_{0}<E_{k},\widetilde{E}_{0}<\widetilde{E}_{p}$
then $\upsilon_{\uparrow\downarrow}<0$. An operator of the
interaction is (within a single molecule the effects of screening
can be neglected)
\begin{equation}\label{3.6}
   \widehat{V}(\textbf{R}_{1},\textbf{R}_{2},\textbf{r}_{1},\textbf{r}_{2})=
   \frac{e^{2}}{|\textbf{R}_{1}-\textbf{r}_{1}|}+\frac{e^{2}}{|\textbf{R}_{1}-\textbf{r}_{2}|}+
   \frac{e^{2}}{|\textbf{R}_{2}-\textbf{r}_{1}|}+\frac{e^{2}}{|\textbf{R}_{2}-\textbf{r}_{2}|}
   -\frac{2e^{2}}{|\textbf{R}_{1}|}-\frac{2e^{2}}{|\textbf{R}_{2}|}.
   \end{equation}
The van der Waals interaction is a high-frequency process because
the interaction is result of virtual transitions between atomic
(molecular) levels. This means that screening of the interaction
must be determined by a high-frequency dielectric function
$\varepsilon_{\infty}$, that is by a plasmon mechanism. However
transition between molecular levels of $\texttt{C}_{60}$, hence
between energy bands of $\texttt{K}_{3}\texttt{C}_{60}$, takes
place. This means that plasmons can not exist with such energies.
The frequency of the transitions between the levels of a helium
atom is even lager: $\sim 20\texttt{eV}$. Hence screening of the
van der Waals interaction by condition electrons is  very
inefficient and it can be neglected.

To calculate the energy of van der Waals interaction if electrons
are in the normal state (\ref{3.1a}) we can use an antisymmetric
wave function:
\begin{eqnarray}\label{3.2a}
    \Phi_{0\uparrow\uparrow}\equiv
    \frac{1}{\sqrt{2}}\left[\Omega_{l=5,\gamma_{1}}(\textbf{R}_{1})\Omega_{l=5,\gamma_{2}}(\textbf{R}_{2})
    -\Omega_{l=5,\gamma_{1}}(\textbf{R}_{2})\Omega_{l=5,\gamma_{2}}(\textbf{R}_{1})\right]\nonumber\\
    \longleftrightarrow
    \frac{1}{\sqrt{2}}\left[\Omega_{l,\gamma'}(\textbf{R}_{1})\Omega_{l=5,\gamma_{2}}(\textbf{R}_{2})
    -\Omega_{l,\gamma'}(\textbf{R}_{2})\Omega_{l=5,\gamma_{2}}(\textbf{R}_{1})\right]\equiv\Phi_{k\uparrow\uparrow}.
\end{eqnarray}
Energy of the van der Waals interaction is
\begin{equation}\label{3.4a}
    \upsilon_{\uparrow\uparrow}=\sum_{k,p}\frac{|\langle
    \Phi_{k\uparrow\uparrow},\Psi_{p}|\widehat{V}|\Psi_{0},\Phi_{0\uparrow\uparrow}\rangle|^{2}}
    {E_{0}+\widetilde{E}_{0}-E_{k}-\widetilde{E}_{p}}.
\end{equation}
It should be noted that due to electroneutrality of a helium atom
and small size of the atom compared to radius of the molecule
$\langle r\rangle=0.31\texttt{A}\ll R=3.55\texttt{A}$ we have that
$\langle 00|V|00\rangle=0$ - the first order process can be
neglected. Moreover we can neglect the exchange processes between
the helium atom and electrons on the molecule's surface, that can
not be done, for example, for atoms $\texttt{Ar}$ and
$\texttt{Xe}$ due to a not small overlap of the atom's wave
functions with the wave functions of electrons on the fullerene
cage \cite{verkh,mad,pott,Ito1}. In this work we neglect the
higher order processes, i.e. non-additivity of van der Waals
interaction between the helium atom, the pair of excess electrons
and the carbon cage of a fullerene molecula.

To estimate $\upsilon$ a fullerene molecule can be considered as a
sphere, that simplifies the calculation of the matrix elements
$\langle kp|V|00\rangle$. Then the wave functions on the
fullerene's surface $\Omega_{l,\gamma}$ are spherical wave
functions $Y_{l,m}$, where $l$ and $m$ is orbital quantum number
and magnetic quantum number accordingly,
$|\textbf{R}_{1}|=|\textbf{R}_{2}|=R$. Energy of each level is
\begin{equation}\label{3.7}
    E_{l}=\frac{\hbar^{2}l(l+1)}{2m_{e}R^{2}}.
\end{equation}
In a ground state $l=5$ each state is degenerated in $m=-l\ldots
l$. The wave function of a Cooper pair in the ground state is
\begin{equation}\label{3.8}
Y_{5,m}(\textbf{R}_{1})Y_{5,m}(\textbf{R}_{2}),
\end{equation}
and the wave function of some a excited state is:
\begin{equation}\label{3.9}
\frac{1}{\sqrt{2}}\left[Y_{l',m'}(\textbf{R}_{1})Y_{5,m}(\textbf{R}_{2})+Y_{5,m}(\textbf{R}_{1})Y_{l',m'}(\textbf{R}_{2})\right],
\end{equation}
where $l'>5$, $m'=-l'\ldots l'$. The antisymmetric wave functions
are
\begin{eqnarray}\label{3.9a}
\frac{1}{\sqrt{2}}\left[Y_{5,m_{1}}(\textbf{R}_{1})Y_{5,m_{2}}(\textbf{R}_{2})-Y_{5,m_{1}}(\textbf{R}_{2})Y_{5,m_{2}}(\textbf{R}_{1})\right]
\longleftrightarrow\frac{1}{\sqrt{2}}\left[Y_{l',m'}(\textbf{R}_{1})Y_{5,m_{2}}(\textbf{R}_{2})-Y_{l',m'}(\textbf{R}_{2})Y_{5,m_{2}}(\textbf{R}_{1})\right].
\end{eqnarray}
It should be noted that in this case the energies of van der Waals
interaction (\ref{3.4}) and (\ref{3.4a}) are functions of the
magnetic quantum number: $\upsilon_{\uparrow\downarrow}(m)$,
$\upsilon_{\uparrow\uparrow}(m)$. Below we will calculate
$\upsilon$ for each $m$. The Coulomb potential is convenient to be
expanded in spherical harmonics:
\begin{equation}\label{3.13}
\frac{1}{|\textbf{R}-\textbf{r}|}=\left\{
\begin{array}{cc}
  \frac{4\pi}{R}\sum_{L,M}\frac{1}{2L+1}\left(\frac{r}{R}\right)^{L}Y_{L,M}^{+}(\theta,\varphi)Y_{L,M}(\Theta,\Phi), & r<R \\
  \\
  \frac{4\pi}{r}\sum_{L,M}\frac{1}{2L+1}\left(\frac{R}{r}\right)^{L}Y_{L,M}^{+}(\theta,\varphi)Y_{L,M}(\Theta,\Phi), & r>R \\
\end{array}\right\},
\end{equation}
where $M=-L, -L+1\ldots L$. Since $\langle
r\rangle=0.31\texttt{A}\ll R=3.55\texttt{A}$ then the expansion at
$r>R$ can be omitted. The matrix element of the interaction can be
represented as
\begin{equation}\label{3.14}
    \langle \Phi_{k\uparrow\downarrow},\Psi_{p}|\widehat{V}|\Psi_{0},\Phi_{0\uparrow\downarrow}\rangle=2e^{2}\sum_{L,M}ABC,
\end{equation}
where
\begin{eqnarray}\label{3.15}
A&=&\frac{4\pi}{R}\frac{1}{2L+1}\int_{0}^{\infty}
f_{n,\widetilde{l}}^{+}f_{0,0}r^{2}\left(\frac{r}{R}\right)^{L}dr\nonumber\\
B&=&\int Y_{\widetilde{l},\widetilde{m}}^{+}Y_{0,0}Y_{L,M}^{+}\sin\theta d\theta d\varphi\nonumber\\
C&=&\int Y_{l',m'}^{+}Y_{5,m}Y_{L,M}\sin\Theta d\Theta
d\Phi\nonumber
\end{eqnarray}
Transitions to states $l'=6$ for a Cooper pair and to states
$2p,3p,4d$ for a helium atom when the magnetic quantum numbers
change as $m'-m=\widetilde{m}=0,\pm 1$ give the largest
contribution to Eq.(\ref{3.14}) . $E_{5,m}-E_{6,m'}\approx
3.6\texttt{eV}$,
$\widetilde{E}_{0,0,0}-\widetilde{E}_{n,\widetilde{l},\widetilde{m}}\approx
21\texttt{eV}$. Calculation shows that $\upsilon$ is almost
independent of $m$:
\begin{equation}\label{3.16}
    \upsilon_{\uparrow\downarrow}(m=0\ldots\pm 5)\approx -80\texttt{K},
\end{equation}
that indicates the correctness and usability of the spherical wave
functions instead of the functions (\ref{3.2}) to estimate the
potential $\upsilon$. Moreover the matrix element $\langle
\Phi_{k\uparrow\uparrow},\Psi_{p}|\widehat{V}|\Psi_{0},\Phi_{0\uparrow\uparrow}\rangle$
does not depend on $m_{2}$. It is easy to show that each matrix
element $\langle
\Phi_{[l,m,l',m']\uparrow\downarrow},\Psi_{p}|\widehat{V}|\Psi_{0},\Phi_{[l,m]\uparrow\downarrow}\rangle$
corresponds to the element $\langle
\Phi_{[l,m_{2},l',m']\uparrow\uparrow},\Psi_{p}|\widehat{V}|\Psi_{0},\Phi_{[l,m_{1},m_{2}]\uparrow\uparrow}\rangle$
where $l'=l'$, $m'-m=m'-m_{1}$. Then we can find that
\begin{equation}\label{3.17}
 \langle \Phi_{k\uparrow\downarrow},\Psi_{p}|\widehat{V}|\Psi_{0},\Phi_{0\uparrow\downarrow}\rangle=
 \sqrt{2} \langle
 \Phi_{k\uparrow\uparrow},\Psi_{p}|\widehat{V}|\Psi_{0},\Phi_{0\uparrow\uparrow}\rangle
\end{equation}
for the corresponding matrix elements. Hence we have
\begin{equation}\label{3.18}
    \upsilon_{\uparrow\downarrow}=2\upsilon_{\uparrow\uparrow}
\end{equation}
Thus a fullerene molecule with excess electrons has a lower energy
if the electrons are in the paired state (\ref{3.1}) than energy
if the electrons are in the state according to Hund's rule
(\ref{3.1a}). For each pair the energy gain is
$\upsilon=-40\texttt{K}$ - Eq.(\ref{3.2b}). The reason of the
relation (\ref{3.17}) consists in the fact that if electrons are
in in different states (with $m_{1}\neq m_{2}$ - Eq.(\ref{3.9a}))
then the probability amplitude is parted into two mutually
orthogonal parts with the weights $1/\sqrt{2}$ each. It should be
noted that for symmetric combination with $m_{1}\neq m_{2}$ - a
plus sign instead of a minus sign in Eq.(\ref{3.9a}) the result is
the same.

\begin{figure}[ht]
\includegraphics[width=8.5cm]{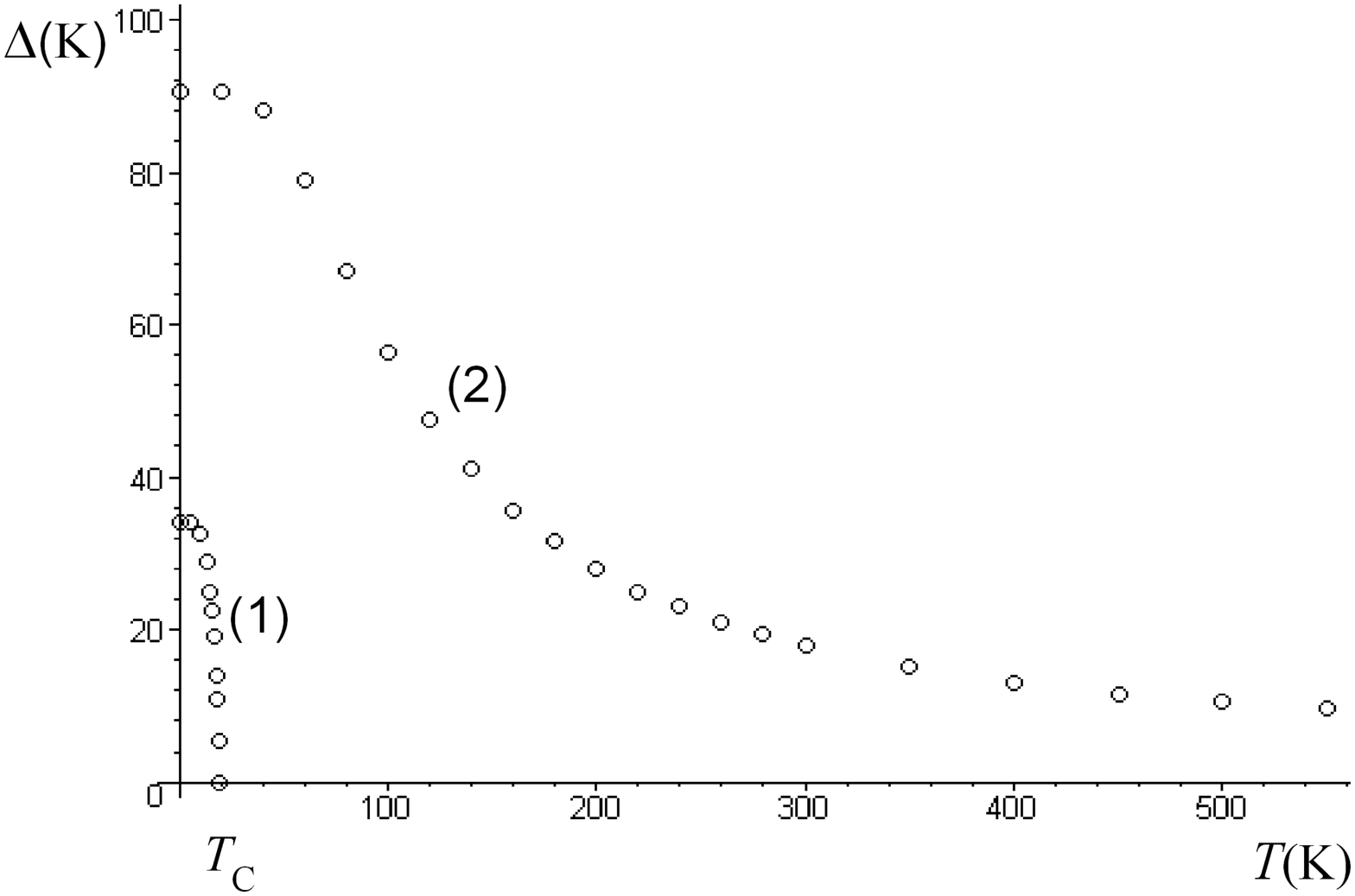}
\caption{Energy gaps $\Delta(T)$ as solution of Eq.(\ref{2.12})
for $\texttt{K}_{3}\texttt{C}_{60}$ with critical temperature
$T_{\texttt{C}}=19.3\texttt{K}$ - a curve $\{1\}$, as solution of
Eq.(\ref{2.13}) for $\texttt{K}_{3}\texttt{He}@\texttt{C}_{60}$
with the external pair potential $\upsilon= -40\texttt{K}$ - a
curve $\{2\}$.} \label{Fig3}
\end{figure}

To estimate the energy gap $\Delta$ we can use the continuum
approximation (\ref{2.13}) instead the Habbard hamiltonian
(\ref{3.1}). The characteristic frequency of intramolecular
$H_{g}$ oscillations can be taken as the effective frequency
$\omega$. Let the frequency is $\omega/W=0.1$ where
$W=0.5\texttt{eV}$ is the electron bandwidth. The effective
coupling constant $g$ must be such that the critical temperature
calculated with Eq.(\ref{2.12}) is equal to critical temperature
$T_{\texttt{c}}=19.3\texttt{K}$ of
$\texttt{K}_{3}\texttt{C}_{60}$, then we have $g=0.283$. The
functions $\Delta(T)$ calculated with Eqs.(\ref{2.12},\ref{2.13})
are shown in Fig.\ref{Fig3}. We can see for $\upsilon<0$ the gap
tends to zero asymptotically with increasing temperature. Thus the
critical temperature is equal to infinity, in practice it is
limited by the melting point of the substance.

\section{Conclusion.}\label{concl}

In this work a model of a hypothetical room-temperature
superconductor has been proposed. Our idea is based on the fact
that the phase independent external pair potential can be added in
BCS Hamiltonian - Eq.(\ref{2.1}). This field acts on a Cooper pair
changing its energy relative to uncoupled state of the electrons.
In a case of increasing of Cooper pair's energy by the external
pair field a suppression of superconductivity takes place. In a
case of decreasing of Cooper pair's energy by the field the energy
gap tends to zero asymptotically with increasing temperature -
Fig.\ref{Fig1}. Thus the ratio between the gap and the critical
temperature is $2\Delta/T_{\texttt{c}}=0$ instead of a finite
value in BCS theory. For practical realization of this model we
propose a hypothetical superconductor on the basis of alkali-doped
fullerenes $\texttt{K}_{3}\texttt{C}_{60}$ or
$\texttt{Rb}_{3}\texttt{C}_{60}$ with the use of endohedral
structures $\texttt{He}@\texttt{C}_{60}$, where a helium atom is
in the center of each fullerene molecule. According to
\cite{han1,lam,lam2,han2} in alkali-doped fullerene Cooper pairs
are formed on surface of the fullerene molecules due
electron-vibron interaction and suppression of hopping between
molecules by one-cite Coulomb interaction. In an endohedral
fullerene the noble gas atom interacts with a carbon cage by van
der Waals force. The van der Waals interaction depends on a state
of excess electrons on surface of the molecule. We have shown that
energy of the molecule if the excess electrons on its surface are
in the paired state (\ref{3.1}), when two electrons are in a state
with the same quantum numbers, is lower than the energy if the
electrons are in the normal state (\ref{3.1a}), when the electrons
are in a state with different quantum numbers and maximal spin. It
makes more energetically favorable to place electrons in the state
with the same quantum numbers $m$, that resists the destruction of
local pairs by the hopping between molecules. Thus difference of
the energies of the van der Waals interaction plays a role of the
external pair potential $\upsilon<0$. We have calculated the
temperature dependence of the energy gap for a hypothetical
material $\texttt{K}_{3}\texttt{He}@\texttt{C}_{60}$ using the
continuum approximation (\ref{2.13}) with an effective
electron-phonon interaction - Fig.\ref{Fig3}. In this material the
superconducting phase exists at any temperature unlike the pure
system $\texttt{K}_{3}\texttt{C}_{60}$ where the phase is limited
by the finite critical temperature.

In connection with the obtained results it should be noted that in
works \cite{Ito2,Ito3} it had been reported about the synthesis of
the first endohedral fullerene superconductors
$\texttt{A}_{3}\texttt{Ar}@\texttt{C}_{60}$ having critical
temperatures on $2-3$ kelvins less than critical temperatures of
the pure materials $\texttt{A}_{3}\texttt{C}_{60}$. The van der
Waals radius of $\texttt{Ar}$ is slightly more than radius of the
inner cavity in center of a fullerene molecule. In this case an
overlap of the argon atom's wave functions with the wave functions
of electrons on the fullerene cage takes place, hence the exchange
interaction plays a role. The radii of $\texttt{Kr}$ and
$\texttt{Xe}$ are much larger hence role of the exchange
interaction is more significant. Thus the influence of
$\texttt{Ar},\texttt{Kr},\texttt{Xe}$ requires special
consideration that goes beyond the present work.



\acknowledgments


The authors would like to thank Dr. G. E. Volovik for helpful
remarks on the work.

\end{document}